\documentclass[preprint]{vgtc}               % preprint
\graphicspath{{figures/}{pictures/}{images/}{./}} % where to search for the images

\usepackage{times}                     % we use Times as the main font
\usepackage{tabu}                      % only used for the table example
\usepackage{booktabs}                  % only used for the table example
\usepackage{lipsum}                    % used to generate placeholder text
\usepackage{mwe}                       % used to generate placeholder figures

\usepackage{mathptmx}                  % use matching math font

\usepackage[shortlabels]{enumitem}

\newcommand{\figarrow}[1]{\protect\tikz[baseline=-0.75ex]\protect\draw[->,very thick,#1](0,-0.5ex)--(0.5em,0.25ex);}

\onlineid{0}

\vgtccategory{Research}

\vgtcinsertpkg

\ieeedoi{10.1109/VIS55277.2024.00010}

\usepackage{tikz}
\title{DaVE - A Curated Database of Visualization Examples}

\author{Jens Koenen*\thanks{equal contributions}\thanks{ email: koenen@vis.rwth-aachen.de}\\ %
     {\scriptsize \centering RWTH Aachen University, Germany} %
\and Marvin Petersen*\\ %
     {\scriptsize \centering RPTU Kaiserslautern-Landau, Germany}
\and Christoph Garth\\ %
     {\scriptsize \centering RPTU Kaiserslautern-Landau, Germany}
\and Tim Gerrits\\ %
        {\scriptsize \centering RWTH Aachen University, Germany}}
 
\teaser{
  \centering
  \includegraphics[alt={
Figure showing a network of images and icons connected by black, red, and blue arrows.
Black arrows describe the workflow of a domain scientist collecting data using an HPC system.
Starting from a potential user in the center of the figure, a black arrow points to an HPC cluster.
From there, another black arrow points to simulation data, while the last black arrow points back to the user.
After collecting the data, the domain scientist would follow the red arrows.
Starting from the user, a red arrow points to several screenshots of the DaVE website.
From there, the next red arrow points to an image of a glyph-based vector field visualization, while the last red arrow points back to the user.
After finding a suitable visualization technique on the DaVE website, the user would follow the blue arrows to deploy the selected technique on an HPC system.
A chain of blue arrows connects the user, screenshots of the DaVE website, several script files, and a three-dimensional cube containing script files representing containerized code.
In addition to the blue arrow coming from the script files, another blue arrow points from the simulation data to the containerized code.
Finally, a chain of blue arrows leads from the containerized code, over the HPC system and the image of the vector field visualization, back to the user.
  },width=0.91\linewidth]{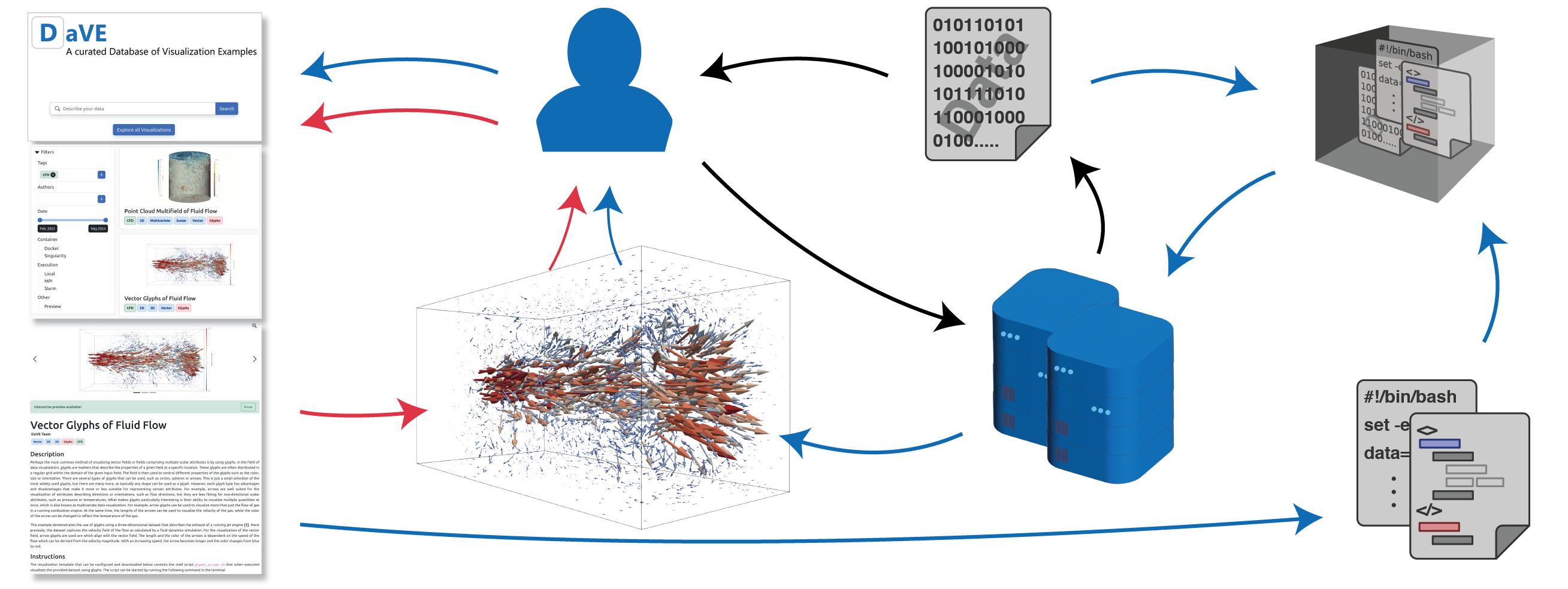}
  \caption{
Schematic overview of a potential workflow with DaVE: 
A domain scientist acquires data, e.g., by running a simulation on an HPC cluster (\figarrow{black}).
Based on this data, they enter search terms into DaVE, which provides a list of appropriate visualization techniques from its curated and extensible database for different data types and domains (\figarrow{red}).
The user can freely explore the examples until a suitable visualization is chosen.
Each example comes with a description, explanation of techniques, additional references, and adaptable containerized code.
This allows the visualization to be calculated on suitable hardware using the respective data without the potentially cumbersome setup process (\figarrow{blue}).}
  \label{fig:teaser}
}

\abstract{
Visualization, from simple line plots to complex high-dimensional visual analysis systems, has established itself throughout numerous domains to explore, analyze, and evaluate data.
Applying such visualizations in the context of simulation science where High-Performance Computing (HPC) produces ever-growing amounts of data that is more complex, potentially multidimensional, and multi-modal, takes up resources and a high level of technological experience often not available to domain experts.
In this work, we present \textit{DaVE} - a curated database of visualization examples, which aims to provide state-of-the-art and advanced visualization methods that arise in the context of HPC applications.
Based on domain- or data-specific descriptors entered by the user, DaVE provides a list of appropriate visualization techniques, each accompanied by descriptions, examples, references, and resources.
Sample code, adaptable container templates, and recipes for easy integration in HPC applications can be downloaded for easy access to high-fidelity visualizations.
While the database is currently filled with a limited number of entries based on a broad evaluation of needs and challenges of current HPC users, DaVE is designed to be easily extended by experts from both the visualization and HPC communities.
} % end of abstract


\begin{document}


\firstsection{Introduction}

\maketitle
%% \section{Introduction} %for journal use above \firstsection{..} instead
Nowadays, developers of research software and simulations in various scientific domains must possess expertise in diverse fields.
This includes not only domain-specific knowledge of models as well as their physical and mathematical background.
Developers are expected to write highly optimized code and find appropriate ways to analyze their results.
The steady advancements in those areas drastically increase the complexity of data and require adapted and new visualization techniques, both on a conceptual level (novel visualization techniques) and on a technological level (efficient HPC implementation of visualization techniques).
While such approaches are being actively developed and discussed within the visualization community, domain experts and HPC users often lack the resources to follow and keep up-to-date with these developments.
This can lead to undesired effects such as a high overhead introduced by finding, learning, and applying approaches from scratch, or being unaware of new and potentially very helpful visualization techniques.
Moreover, on the technological side, while implementations of modern visualization algorithms suitable for HPC use exist, incorporating these into complex HPC application stacks requires substantial expertise and an investment of additional engineering effort.
Thus, while domain scientists see the need and are very open to using more advanced visualization techniques,
they often lack resources and references, to use them.\\

In this work, we present DaVE\footnote{\url{https://dave-vis.com}} - a curated database of visualization examples targeted at novice and experienced HPC users alike, that allows for easy access to state-of-the-art and advanced visualization methods.
The database is accessible through a modern and fast web interface and includes detailed descriptions, sample media, references, and example implementations tailored to HPC projects which allows for easy integration in existing projects and application stacks.
Users receive guidance not only for approaches they know and want to get more information on, but also recommendations on best practices or alternative visualization methods.
The underlying system is developed in such a way that curators and users can easily extend the database with their own information and examples to establish a platform of best practices, current developments, and support. 
As such, DaVE aims to function as a focal point and community resource for visualization within the national HPC community and beyond.
The database has the potential to develop into a general resource for visualization regardless of application.
%%
%To the best of our knowledge, no similar service or resource exists to date.

\section{Background and Related Work}
Visualization has become an integral part of understanding, analyzing, and communicating complex data in virtually all scientific domains.
Therefore, many data management tools such as Microsoft Excel, Matlab, or Tableau, already provide built-in visualization options for some data types, such as data tables, that make it easy for non-expert users to achieve fast and easy visualizations.
Several projects have emerged in the past that aim to provide easy access to adaptable visualizations for similar or more complex data, such as the D3 library~\cite{bostock2011d3} or Vega-Lite~\cite{satyanarayan2016vega}.
A significant body of work exists on so-called content-based recommendation~\cite{oppermann2020vizcommender, kaur2017review, zeng2021evaluation, kubernatova2018knowledge}, where systems try to automatically find the best-fitting visualization given a set of data.
Naturally, this includes approaches where recent machine learning technologies are leveraged~\cite{hu2019vizml,qian2021learning}, specific tasks are analyzed~\cite{shen2021taskvis}, or user behavior is evaluated~\cite{gotz2009behavior, epperson2022leveraging}.
However, most of these are limited to 2D plots.

Simulations that model physical phenomena often leverage high-performance computing infrastructure for, e.g., efficient computation of computational fluid dynamic (CFD) models or force propagation in complex materials.
Such data typically describes spatial, multimodal 3D data that might also be time-dependent.
While some of that data can be summarized and described by statistical data, visual exploration usually requires efficient interaction with 3D geometries such as direct volume renderings or derived features of often very large data quantities.
Besides efficient handling and processing, such data size might even require in-situ processing or rendering which further increases the complexity of visualization.
Here, too, a number of tools and libraries exist, such as VTK~\cite{vtkBook} or ParaView~\cite{ahrens200536}, which provide a multitude of options such as filters and renderers.
In the context of in-situ or in-transit visualizations, distinct visualization libraries such as Catalyst~\cite{catalyst} or Ascent~\cite{ascent} are available.
Learning these tools or integrating them into existing simulation code, however, requires time and knowledge.
For libraries, such as D3, it has been shown that providing a database of examples and templates can considerably reduce a user's effort to design visualizations \cite{bako2022understanding, bako2022streamlining}.
VTK, too, has example code available on a dedicated website, and approaches such as EZ-ISAV~\cite{ezisav} seek to make the integration of visualization in complex software stacks easier by providing containerized solutions.
However, to the best of our knowledge, there is no publicly available resource that allows non-visualization experts to easily find suitable techniques primarily for HPC-based simulation data.

\section{DaVE}\label{sec:DaVE}
\begin{figure}[ht!]
    \centering
    \includegraphics[alt={
Figure showing the gallery of visualization examples that DaVE present to the user.
At the top of the page is a navigation bar that contains links to other pages of DaVE, a text field where the user can enter search queries, and a drop-down menu that allows the user to control how the examples are sorted.
On the left side of the page is a box with additional filters that the user can use to further narrow the search.
In addition to limiting the search to specific authors or a specific time period when the examples were added to the database, the user can also specify which tags are required to be present in the search result.
Furthermore, the user can use checkboxes to control, for example, whether the search results should provide an interactive preview or be executable on a cluster using MPI or Slurm.
The heart of the page, however, is the gallery of visualization examples, which is organized in a three-column grid.
Each visualization example matching the user's search parameters is represented in the gallery by a box containing an image, the name of the example, and the tags associated with it.
For the given search query shown in the figure, the user is presented with 9 visualization techniques, ranging from simple line-based visualization to complex volume or iso-surface visualization.
    },width=0.95\linewidth]{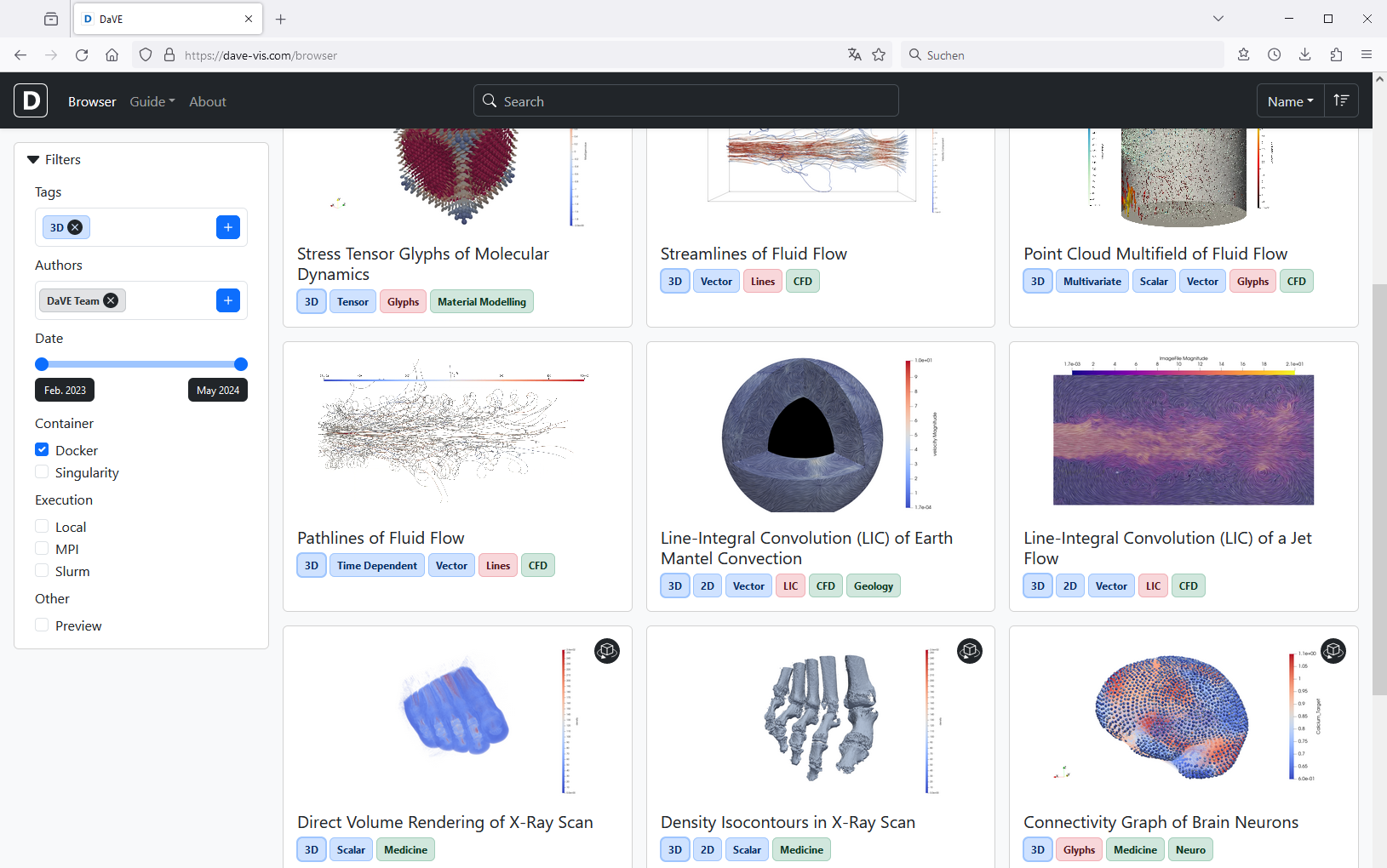}
    \caption{Through the adaptive web interface, DaVE provides a gallery of visualization examples that can be searched and filtered through tags and requirements.}
    \label{fig:overview}
\end{figure}
For us, the need for an easily accessible resource for HPC users to discover visualization examples became apparent after collaboration with several simulation scientists within a large national research project.
An internal user evaluation~\cite{tim_gerrits_2022_7715663} revealed that HPC users across domains all appreciated more sophisticated state-of-the-art visualization techniques, such as in-situ/in-transit capabilities, visual analysis and exploration systems, and immersive visualizations.
However, all lacked resources, guidance, and necessary knowledge to find and integrate suitable techniques.
Recognizing this gap, we believe that a curated database of such techniques and examples can be a way to address these issues.
Users also expressed concerns that testing and integrating such techniques requires substantial effort as well as changes to existing and often complicated code bases.
After several discussions at workshops, conferences, and personal meetings, we identified three essential requirements for a successful resource.
These can be summarized as follows:
\begin{enumerate}[label={\bfseries R\arabic*}]
    \item A non-visualization expert-friendly system to find domain- or data-specific visualization techniques
    \item Easy integration of visualization examples into existing HPC infrastructure
    \item Easy extensibility of the database with new and relevant information
\end{enumerate}

Based on these, we decided to create DaVE - a curated database of visualization examples.
%%
%It provides a searchable database accessible via the web interface, while each example consists of descriptions and resources, including adaptable containers to test examples on different hardware configurations.
%%
In the following, we will describe and justify the individual components and functionalities.

\subsection{Web Platform for Visualization Exploration}\label{sec:web}
\begin{figure}[h!]
    \centering
    \includegraphics[alt={
Figure showing the web page that the user sees when they select and click on a visualization example in the example gallery.
In this case, a glyph-based visualization of a three-dimensional vector field entitled "Vector Glyphs of Fluid Flow" has been selected.
At the top left of the page is a collection of images that illustrate the vector field of a fluid flow using arrows.
The currently visible image shows arrows, ranging from blue over orange to red, placed inside a box and pointing primarily to the right.
An arrow-shaped button is placed at the left and right edge of the image, allowing the user to scroll to the next or previous image.
Below the preview images is a green box with the text "Preview available!" and a button labeled "Show" in it.
At the bottom of the page and below the green box is additional information such as the title of the visualization example, the author, and the tags associated with the example.
This example was created by the DaVE team and has the blue tags "Vector", "2D", "3D", the red tag "Glyphs", and the green tag "CFD".
On the right side of the web page is an outline that gives an overview of the content of the page.
According to this, the web page also contains a description, instructions on how to use the example, limitations of the example, references, and additional resources.
    },width=0.98\linewidth]{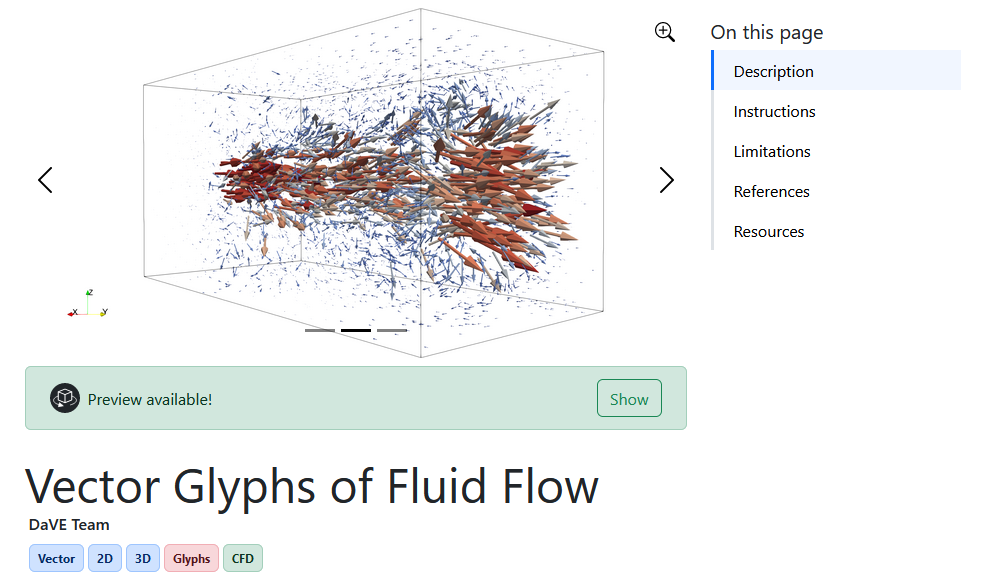}
    \caption{Tags describing data type, visualization technique, and domain of the example are included in the search.
    Example images, optional interactive previews, and text descriptions provide further information on the technique and usage of provided resources.}
    \label{fig:example-page}
\end{figure}

\begin{figure}[hb!]
    \centering
    \includegraphics[alt={
The figure shows the interactive preview of a visualization example that allows the user to explore and test the visualization directly on the DaVE website.
In the preview, a three-dimensional scalar field describing the density variation within a human foot is visualized by three slices that cut through the data set.
The slices show the empty space around the foot in dark blue, the tissue of the foot with relatively low density in light blue, and the bones within the foot with very high density in red.
On the right side of the preview is a scale from dark blue to light blue to light red to dark red that indicates the mapping from density values to the shown color range.
Dark blue corresponds to a density of zero, while dark red corresponds to a density of 250.
    },width=0.98\linewidth]{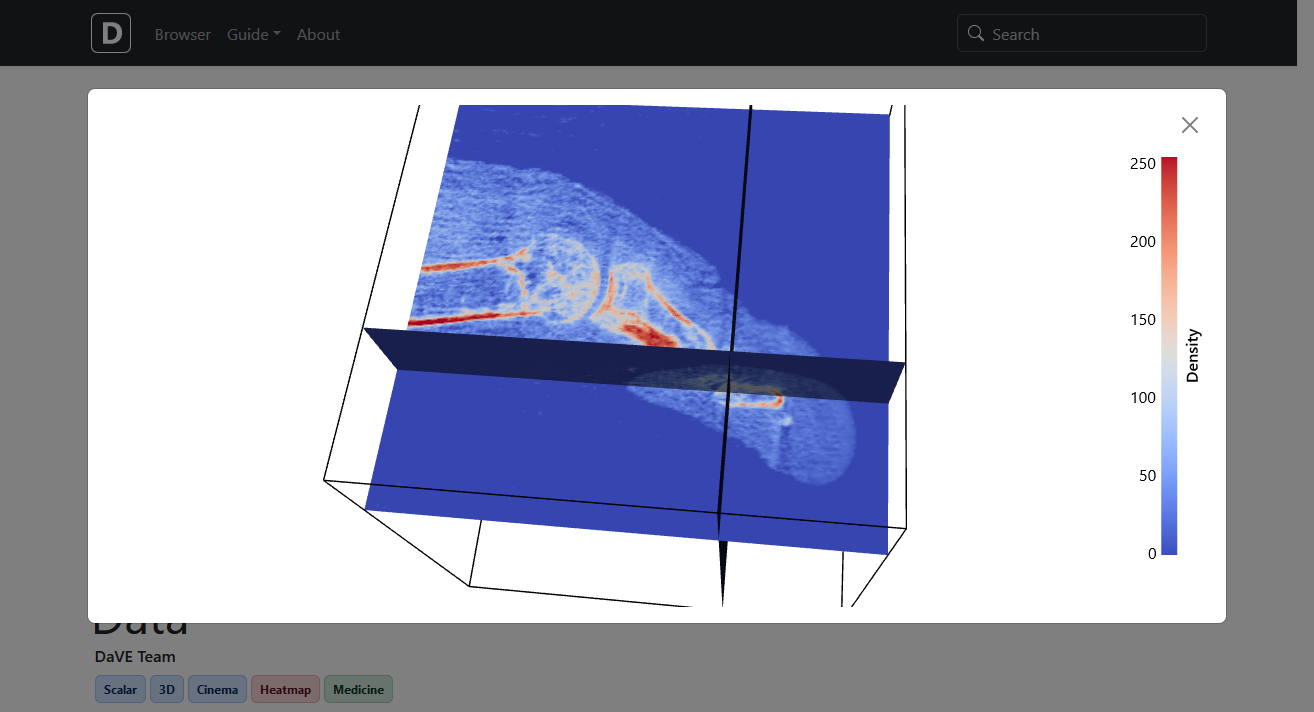}
    \caption[DaVE]{DaVE supports interactive examples on the websites through VTK.js~\footnotemark, allowing users to directly test visualizations.}
    \label{fig:preview}
\end{figure}

DaVE's main point of interaction is a responsive website.
Users are prompted to enter a search term that describes their data or needs.
This search is aided by suggestions of suitable keywords.
From there the user is provided with a gallery overview of the visualization examples that fit the search description. 
For each example, a preview image and tags for categorization are displayed.
In a sidebar, these items can be further filtered by tags, date, author, and other parameters (see \autoref{fig:overview}).
Navigating to one of the example pages (see \autoref{fig:example-page}) shows an introductory text, covering the visualization methods, its use cases, and basic functionality.
Further, short instructions and materials for reproducing the example are provided.
Also, examples may offer an interactive preview of the visualization (see \autoref{fig:preview}). 
With this design, users directly interact with the system, to help them find solutions for their visualization problem before acquainting them with the needed technology (\textbf{R1}). %and only then being able to check whether it suits their needs or knowing whether what they want to achieve is even possible.
A complete package for executing the example can be configured and downloaded, which is executed by calling a single script.
Configuration options include the path for own datasets, the usage of MPI and Slurm, and the container technology, either Docker or Singularity/Apptainer~\cite{singularity}, that is available on the target machine.
Dedicated guide pages contain general information about DaVE, the structure of an example, and instructions on how to adapt existing or create custom examples in general. 
These pages also include references to the documentation of used software for further customization. 
Additionally, the guide pages specify how adapted and new examples can be added to DaVE, simplifying the contribution process (\textbf{R3}). 
Providing the executable templates and information allows for exploratory tinkering, and acquiring knowledge about the underlying technologies in the process.
Each example has the option to open an issue on the project github that allows to provide feedback or questions if clarifications are needed.

\footnotetext{\url{https://github.com/Kitware/vtk-js}}

\subsection{Integration Through Containerized Examples}
To be able to allow users working on a large variety of hardware configurations, from personal computers to large high-performance computing clusters, to test and explore the examples, DaVE provides containerized environments.
This has the following advantages:
A containerized environment can easily be reused for multiple examples, reducing the download time if multiple similar examples are explored.
Furthermore, these containers offer high portability.
Most examples that are currently available via DaVE can be executed locally, with MPI, or via Slurm on HPC clusters (\textbf{R2}). 
Additionally, extending containerized environments is straightforward. 
The containers can be extended by using them as a base for a new custom image or users can change the build recipe in the original directly (\textbf{R3}).
However, and most importantly, the provided containers do not have to be extended or changed at all to enable using the given examples. 
This allows users from different domains with varying background knowledge to explore visualization in HPC environments from fully functional examples and start from there instead of having to build a reference use case themselves (\textbf{R1}).

With respect to container technologies, DaVE relies on Docker for creating containers.
The images can be uploaded to a container image library, like Docker Hub.
From there, the executable package of an example in DaVE retrieves the image either via Docker itself or with singularity.
In order to build maintainable environments inside the containers, the required software and its dependencies are configured and built with spack~\cite{spack}.
It is a flexible package manager that streamlines the process of managing software dependencies, making it easier to install, configure, and maintain projects, which, in turn, reduces time and effort for creating examples.

\subsection{Extending DaVE}
DaVE supports the easy addition of new techniques and examples.
Entries of the database are built upon a straightforward file and folder structure.
Each example is defined by a folder on DaVE's webspace which contains text files including textural explanations, as well as descriptive tags and meta information.
Subfolders contain further downloadable resources and files describing options for container creation.
Adding a new example is as simple as adding a new folder at the respective location.
This can either be done by the original authors by request, or by submitting a merge request on the project github.
This allows and encourages an active community to maintain the quality and correctness of the entries.

As mentioned in Section \ref{sec:web}, a dedicated page on DaVE further provides all necessary information, instructions, and downloadable templates to create own examples or alter existing containers to work with additional software.
There are several levels of complexity for adding new entries, allowing users with different backgrounds and varying familiarity with the technologies used to contribute (\textbf{R1},\textbf{R3}). 
The simplest option is to adapt an existing example to new data.
This requires no changes in the containerized environment and only minor changes in the visualization procedure. 
Creating custom visualizations is the next option.
Again no changes in the environment are needed, but more knowledge about the visualization software is required.
In the case of ParaView, users need to know how to create and manipulate ParaView's Python trace files.
If no available container provides the necessary tools for a new example, existing containers can be extended either via docker or reconfigured and built with spack.
Since managing software for use with HPC infrastructure can be a daunting task for non-experts, spack provides easy configuration capabilities for the existing containers.
This setup allows users to acquire knowledge on specific areas of visualization in HPC in a step-by-step manner. 
Our provided examples serve as a starting point to dive deeper into the respective techniques, like Docker, spack, or ParaView.
For the greatest flexibility, fully custom docker images can be used.
As such, we not only hope to establish DaVE as a standard resource for HPC users but also to encourage active feedback and development by the respective communities.

\section{Evaluation and Expert Feedback}
To get an indication, of whether our current design of DaVE fulfills the requirements outlined in Section~\ref{sec:DaVE}, we invited five simulation scientists from different research institutions working on various domains to evaluate DaVE.
None of the participants had been involved in the project before but all were frequent users of HPC infrastructure with research experience ranging from 3 to 10 years.
Domains of expertise included computational fluid dynamics, material design, artificial intelligence, and energy conversion.
All but one had worked with containers before and considered themselves familiar with containerized code.
Participants were asked to freely explore DaVE on their own devices, either under observation at their workplace or via Zoom recording.
Sessions typically lasted between 15 and 40 minutes, during which users were asked to verbalize their thoughts and actions using a think-aloud strategy~\cite{lewis1982using}.
We further prepared a short questionnaire based on the System Usability Score (SUS)~\cite{brooke1996sus} and additional questions that participants received after they had finished their exploration of DaVE.
At the time of the evaluation, the database contained 17 curated examples across various domains added by the authors.
Participants used their desktop work environment and standard modern web browsers such as Firefox, Chrome, or Safari.

Users started on the landing page and understood the core functionality of DaVE from reading the provided overviews and the use of a standard search bar.
The simplicity, modern design, and clear focus on the search functionality were mentioned and appreciated by all.
While one user opted for entering a domain-specific search term in the search bar, others chose to explore all examples.
Entering the gallery, participants quickly understood the tagging/filtering system, added and removed tags to test it out, and mentioned them as an intuitive and very useful method to browse the database.
Users would scan the preview images and example names to look for familiar data before using the tags to narrow down results.
Some users found the limited number of examples within their specific domain unfortunate but viewed this as a temporary limitation that could be overcome through active community contributions.
Once an example was selected, users especially pointed out the usefulness of interactive previews and focused descriptions.
Some suggested more interactive previews, e.g., allowing to apply filters or change color mappings.
After experimenting with different options, users would return to the gallery to look for more examples.
Only one user configured, downloaded, and started a containerized example, but did not need any instructions to do so.
Questions that users had during interaction with DaVE were always answered by either experimenting with features or consulting the provided information on the website.
Minor comments on design decisions as well as various feature ideas were verbalized throughout the process.
After users claimed they were done exploring, all stated they had enjoyed exploring DaVE.
They appreciated the overall idea and effort invested in the system, and recognized its potential, especially with the addition of more examples.

Evaluating the questionnaire, DaVE received a mean system usability score of 83.5, classified as ``\emph{excellent}" according to Bangor et al.~\cite{bangor2009determining}.
Users agreed that DaVE made it \textit{very easy} to \textit{extremely easy} to find domain-data-specific visualization techniques, indicating fulfillment of \textbf{R1}.
All but one reported discovering new visualization approaches, previously unknown to them.
Evaluating \textbf{R2} proved less conclusive: users were \textit{fairly confident} that techniques could be integrated into their own work but reported needing more time to test it.
It was highly appreciated that containerized environments were provided, given participants' familiarity with them, and using ParaView was mentioned to be a reasonable choice, due to its recognition.
However, some hoped for more examples using alternative tools to accommodate other workflows.
Finally, users reported that extending DaVE (\textbf{R3}) appears to be \textit{fairly easy}, but considered themselves more as users of DaVE than as contributors.
They all reported that the level of detail in the information section should suffice.
Yet, one participant saw the potential in DaVE to replace the often fragmented and decentralized collections of code and information within different research groups.

\section{Discussion}

%% Meeting requirements by design - validation - growing DaVE
We carefully designed DaVE to address each of the three requirements comprehensively.
Domain- and data-specific search options and examples directly showcase applicable visualization techniques for non-experts (\textbf{R1}).
Tailored configuration options for containerized environments specifically designed for HPC ensure portability and easy integration into existing infrastructure (\textbf{R2}).
A simple and intuitive example structure, with human-readable configuration files, annotations, and concise instructions for adaptation and contribution, ensures the extensibility of the system (\textbf{R3}).
While we are aware, that the number of participants in the expert interview is limited and provides only initial insights, the results are promising.
Due to the nature of free exploration of DaVE, some features were less thoroughly explored than others.
Ideally, we hope to gather more feedback once the resource is available to a broader audience and has a larger number of examples added, enabling long-term refinement.
Currently, our examples are based on ParaView implementations, but we aim to cover a broader variety of approaches in the future.
The current strategy for establishing DaVE as a general resource for many communities consists of growing and validating the approach inside collaborations in a national project before expanding its scope to use cases from interested groups outside.
%%
%% Maintenance and other challenges
Maintaining such a resource is a multifaceted challenge.
Ensuring continued relevance of information in DaVE involves updating outdated examples and adding new ones for emerging techniques.
With a growing database, this task becomes insurmountable for a small group of maintainers but should be manageable with active community participation.
With increasing user and contributor base, the diversity of target environments is likely to increase as well.
Testing the entries of DaVE against every possible use case seems unpractical, but would greatly improve the relevancy and usefulness of the resource.
Currently, we rely on a community-driven approach, where problems with examples can be reported in the issue tracker of the accompanying repository.
Furthermore, it has to be ensured that the technical infrastructure, including the website and repository, is kept up-to-date and working while highly requested features have to be integrated.
The longevity of this resource is ensured through a national long-term project dedicated to building and maintaining such infrastructure.
%%

% - Keep it relevant for community
% - Longevity/Sustainability is ensured through national long-term project made to build maintained infrastructure
% - who tests?
% - User evaluation is only small
% - making the resource known

\section{Conclusion and Future Work}
In this work, we introduced DaVE - a curated database of visualization examples, which aims to provide domain scientists working on HPC infrastructure a simple way to find, explore, and apply state-of-the-art visualization techniques in their projects.
After evaluating challenges and needs of users within a large national HPC research collaboration, we designed DaVE based on three requirements:
A user-friendly system to browse and find visualizations, adaptable containerized examples for easy integration across diverse hardware configurations, and the option to easily extend the database, such that it can become a community-driven project.
It is accessible via a simple yet easy-to-navigate web interface and provides information and resources based on user search queries.
While the current example size is limited, a preliminary evaluation with expert users yielded promising results, pointing toward the utility of DaVE.
Now that DaVE is publicly available, we hope that it results in more requests from users as well as support from the visualization community to help it grow.

In the future, we will continue working on DaVE by incorporating additional and extending existing examples.
While the expert evaluation already led to a list of potential improvements, we plan to incorporate further feedback from users for possible design and usability enhancements as well as new features:
Automation processes might be used to regularly find and update entries similar to Oppermann et al.~\cite{oppermann2021vizsnippets}.
A search wizard might guide users to more precise intention statements, allowing for more appropriate search terms and technique selections.
We will also explore how machine-learning-based recommendation systems might benefit the system.
Further, we are exploring the idea of developing an API that allows developers to get access to search results and resources, allowing the integration of DaVE results into existing software stacks.

\acknowledgments{
The authors gratefully acknowledge the German Federal Ministry of
Education and Research (BMBF) and the state government of NRW and RP for
supporting this work/project as part of the NHR funding.}

\bibliographystyle{abbrv-doi}

\bibliography{template}
\end{document}